\documentclass[sigconf, table,sigconf]{acmart}
\AtBeginDocument{%
  \providecommand\BibTeX{{%
    \normalfont B\kern-0.5em{\scshape i\kern-0.25em b}\kern-0.8em\TeX}}}

\usepackage{tablefootnote}
\usepackage{multirow}
\usepackage{makecell}
\usepackage{booktabs}
\usepackage{CJKutf8}
\usepackage{amsmath}
\usepackage{array}
\usepackage{booktabs}

\newcommand{\eg}{{e.g.,\ }}

\newcommand{\ie}{{i.e.,\ }}

\newcommand{\ST}{{\#COVID-19 patient seek help [super-topic]\#}}
\newcommand{\totalCount}{8,395\ }

\definecolor{oxfordblue}{rgb}{0.0, 0.13, 0.28}
\definecolor{harvardcrimson}{rgb}{0.79, 0.0, 0.09}
\definecolor{dartmouthgreen}{rgb}{0.05, 0.5, 0.06}
\definecolor{princetonorange}{rgb}{1.0, 0.56, 0.0}
\definecolor{yaleblue}{rgb}{0.06, 0.3, 0.57}
\definecolor{usccardinal}{rgb}{0.6, 0.0, 0.0}
\definecolor{uclablue}{rgb}{0.33, 0.41, 0.58}
\definecolor{msugreen}{rgb}{0.09, 0.27, 0.23}
\definecolor{cornellred}{rgb}{0.7, 0.11, 0.11}
\definecolor{pomegranate}{RGB}{192, 57, 43}
\definecolor{anti-pomegranate}{RGB}{43,178,192}
\definecolor{alizarin}{RGB}{231, 76, 60}
\definecolor{anti-belize}{RGB}{185, 41, 56}
\definecolor{belize}{RGB}{41, 128, 185}
\definecolor{peter}{RGB}{52, 152, 219}
\definecolor{green}{RGB}{22, 160, 133}
\definecolor{anti-green}{RGB}{160,22,118}
\definecolor{turquoise}{RGB}{26, 188, 156}
\definecolor{pumpkin}{RGB}{211, 84, 0}
\definecolor{anti-pumpkin}{RGB}{0,22,211}
\definecolor{carrot}{RGB}{230, 126, 34}
\definecolor{wisteria}{RGB}{142, 68, 173}
\definecolor{anti-wisteria}{RGB}{99,173,68}
\definecolor{amethyst}{RGB}{155, 89, 182}
\definecolor{nephritis}{RGB}{39, 174, 96}
\definecolor{anti-nephritis}{RGB}{174,39,117}

 \newcommand{\ywj}[1]{{\color{black} #1}}

\usepackage[utf8]{inputenc}
\setcopyright{acmcopyright}
\copyrightyear{2022}
\acmYear{2022}
\setcopyright{acmlicensed}\acmConference[CHI '22]{CHI Conference on Human Factors in Computing Systems}{April 29-May 5, 2022}{New Orleans, LA, USA}
\acmBooktitle{CHI Conference on Human Factors in Computing Systems (CHI '22), April 29-May 5, 2022, New Orleans, LA, USA}
\acmPrice{15.00}
\acmDOI{10.1145/3491102.3517591}
\acmISBN{978-1-4503-9157-3/22/04}

\AtBeginMaketitle{%
\let\oldAtBeginDocument\AtBeginDocument%
\renewcommand\AtBeginDocument[1]{#1}
\setcopyright{acmlicensed}
\copyrightyear{2022}
\acmYear{2022}
\acmDOI{10.1145/3491102.3517591}
\acmISBN{978-1-4503-9157-3/22/04}
\let\AtBeginDocument\oldAtBeginDocument%
}

\begin{document}

\title[\resizebox{4.5in}{!}{How to Save Lives with Microblogs? Lessons From the Usage of Weibo for Requests for Medical Assistance During COVID-19}]{How to Save Lives with Microblogs? Lessons From the Usage of Weibo for Requests for Medical Assistance During COVID-19}

\author{Wenjie Yang}
\email{wyangbc@connect.ust.hk}
\affiliation{%
  \institution{The Hong Kong University of Science and Technology}
  \country{Hong Kong, China}
}

\author{Zhiyang Wu}
\email{zwubd@connect.ust.hk}
\affiliation{%
  \institution{The Hong Kong University of Science and Technology}
  \country{Hong Kong, China}
}

\author{Nga Yiu Mok}
\email{nymok@connect.ust.hk}
\affiliation{%
  \institution{The Hong Kong University of Science and Technology}
  \country{Hong Kong, China}
}

\author{Xiaojuan Ma}
\email{mxj@cse.ust.hk}
\affiliation{%
  \institution{The Hong Kong University of Science and Technology}
  \country{Hong Kong, China}
}

\begin{abstract}

During recent crises like COVID-19, microblogging platforms have become popular channels for affected people seeking assistance such as medical supplies and rescue operations from emergency responders and the public.
Despite this common practice, the affordances of microblogging services for help-seeking during crises that needs immediate attention are not well understood. To fill this gap, we analyzed 8K posts from COVID-19 patients or caregivers requesting urgent medical assistance on Weibo, the largest microblogging site in China. Our mixed-methods analyses suggest that existing microblogging functions need to be improved in multiple aspects to sufficiently facilitate help-seeking in emergencies, including capabilities of search and tracking requests, ease of use, and privacy protection. We also find that people tend to stick to certain well-established functions for publishing requests, even after better alternatives emerge. These findings have implications for designing microblogging tools to better support help requesting and responding during crises.

\end{abstract}

\ccsdesc[500]{Human-centered computing~Empirical studies in collaborative and social computing}

\keywords{Crisis informatics, social media, help-seeking, disaster response, COVID-19
}

\maketitle

\section{Introduction}

\begin{figure*}[t]
\includegraphics[width=.8\textwidth]{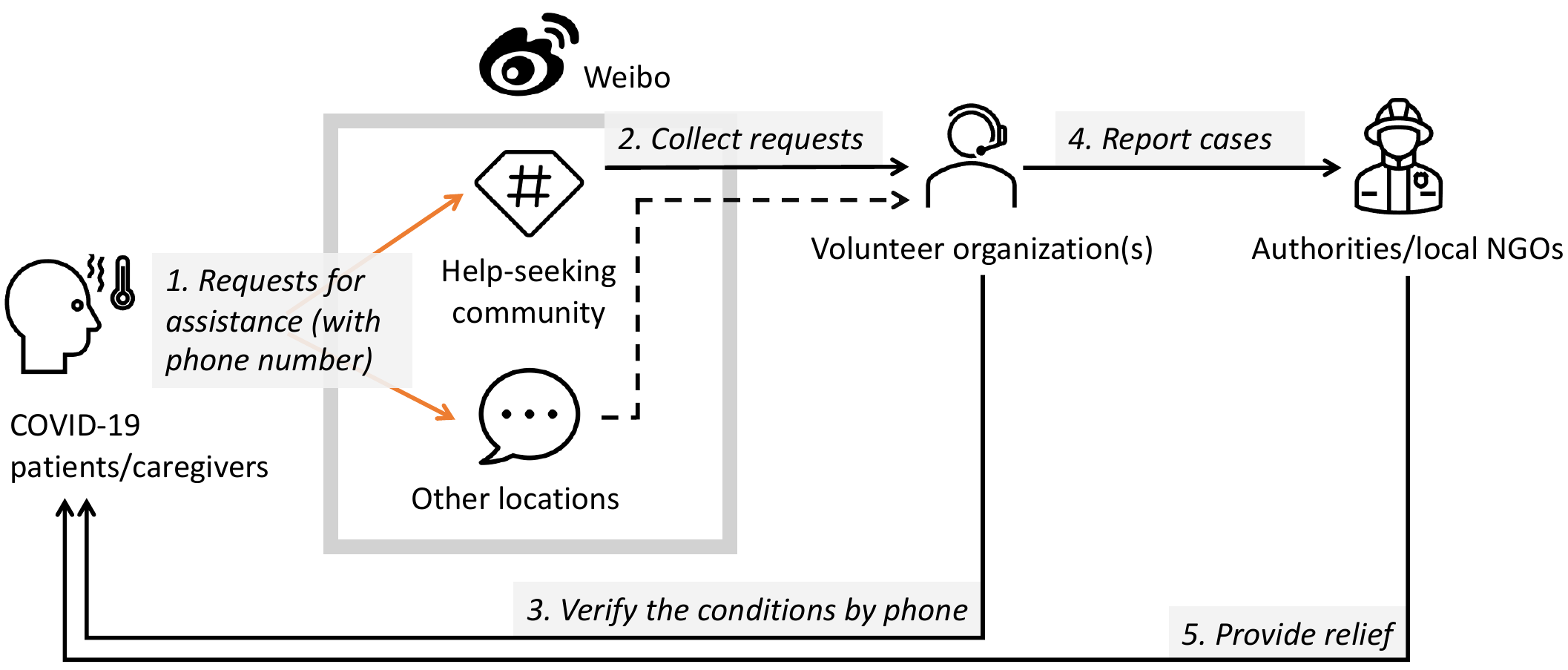}
\centering
\caption{
The process of seeking, providing, and receiving assistance mediated by Weibo, and the information flow between various associated groups. This study primarily focused on how COVID-19 patients or their caregivers use Weibo to seek help, as depicted by the orange arrows above. 
}
\label{fig:bg-info_flow}
\Description{This figure is fully described in the text.}
\end{figure*}

Large-scale crises, such as hurricanes and epidemics, can put a large number of people in danger or even at the risk of death within a short time. Immediate attention and assistance are thus critical to reduce the loss of lives. Unfortunately, traditional emergency services such as crisis hotlines sometimes suffer delays or failures due to various factors such as limited capacity and insufficient manpower \cite{shneiderman2007911, rhodan2017please}. In the absence of adequate support, some people affected by the crises turn to seeking help on their own. For instance, when COVID-19 broke out in Wuhan city, China, many infected citizens without timely medical attention used Weibo, a popular microblogging site in the country, to post their needs such as sick beds and testing kits online and request assistance from the public \cite{huang2020mining}. Volunteer organizations and individuals within and outside the city responded to these requests by providing useful information and coordinating with local authorities and non-governmental organizations (NGOs) that could offer on-site assistance. It was reported that over 3,000 requests on Weibo were submitted to the local government, and at least 318 people obtained assistance \cite{li2020crisis}. 
\ywj{
Similarly, other microblogging platforms, such as Twitter and Facebook, have also been used for help-seeking in various crises around the world \cite{nishikawa2018time, Hiroyuki_FUJISHIRO2018, nishikawa2019analysis, kaufhold2014vernetzte,
qu2011microblogging, bbc_news_2021}.
}

\ywj{
Microblogging is a social media service that allows users to share small pieces of content, such as short sentences and individual images or videos, with large audiences regardless of distance \cite{kaplan2011early}. This Internet-based communication tool facilitates interactive, collaborative, and timely information exchange with high capacity during disasters, providing an alternative to traditional telephone, radio, and television-based communication channels \cite{jaeger2007community, latonero2011emergency, martinez2018twitter}. Over the past decade, it has been used creatively in crisis communication, such as by citizens seeking and sharing situational updates and by emergency responders coordinating relief efforts \cite{cameron2012emergency, bruns2012qldfloods, denef2013social, haunschild2020sticking}. While social media usage patterns during emergencies have been extensively studied in crisis informatics literature, there is limited understanding of how people use microblogging for help-seeking in crisis settings and its affordances for this purpose. 
Filling this knowledge gap could lead to better design of microblogging platforms for streamlining the delivery of help requests (i.e., help-seeking messages or posts concerning disaster victims' lack of essential resources or services  \cite{ullah2021rweetminer}) to emergency responders. Timely attention to help requests can improve emergency teams' situational awareness \cite{vieweg2010microblogging}, support better decisions in planning and executing rescue operations \cite{sun2011rescueme} and resource allocation (e.g., medical care and shelter) \cite{ullah2021rweetminer, zeimpekis2014humanitarian}, and ultimately increase the chance of saving lives.
}

To fill this research gap, we conducted a case study examining the phenomenon of COVID-19 patients seeking medical assistance on Weibo. Our research context lies at the beginning of this public health crisis (around February 2020), when the Chinese government issued a strict lockdown in Wuhan and the city's healthcare system was overstretched. Many infected citizens and suspected patients sought support on Weibo. In view of this demand, Weibo then launched a special community through its super-topic feature (a subreddit-like space for publishing posts related to a designated topic; we will explain it in Section \ref{sec:bg}), dedicated to accommodating the (suspected) COVID patients' needs. Figure \ref{fig:bg-info_flow}
depicts the whole process of how Weibo supports help-related communication, and we focus on the help-seeking behavior in the process and aim to answer two research questions (RQs):
\begin{enumerate}
    \item a) What microblogging functions were exploited by COVID-19 patients or caregivers to request medical assistance during COVID-19 and b) 
    how (in)effective are these functions for this purpose?

    \item How did people use the various microblogging functions to request help?

\end{enumerate}

To answer these RQs, we analyze a dataset of over 100 million epidemic-related posts on Weibo created between January 1 and May 18, 2020 (covering the Wuhan lockdown period). By constructing a machine learning classifier, we identify 8K help requests associated with COVID-19 patients (including suspected cases). 
Using these data, we identify nine types of microblogging functions used in these requests (RQ1),
with the two most popular posting channels being  
the help-seeking super-topic
and regular posting on users' timelines. 
We also recognize some instances of misusing certain microblogging functions. 
We further evaluate these functions based on their efficacy in boosting post diffusion (getting reposted), as well as their limitations in supporting help-seeking in crisis situations, including the capabilities of search and tracking requests, ease of function use, and privacy protection.
For RQ2, we model people's usage of the two main functions mentioned above over time as well as changes in their choice of communication function, if any. Using controlled interrupted time series (CITS) analysis, we assess the possible impact of
the super-topic
appearing on Weibo trending list on help seekers' selection of communication channel. 
The functional limitations we identify
can provide implications for microblogging sites in designing or adapting their functionality to better support urgent requests made by affected groups during emergency.
We find that more people sought help outside the official channel, and they tend to stick to the same platform function to make requests, even after the more efficient channel became available.
These empirical findings contribute to the body of human-computer interaction (HCI) literature on crisis informatics and provide practical implications for emergency responders 
to better communicate with and support crisis-affected people.%

\section{Related Work}

\subsection{Online Help-seeking}
When people face difficulties that they cannot resolve independently, they seek assistance from others by discussing their problems and requesting assistance \cite{gourash1978help}. In recent years, the proliferation of the internet has made the practice of online help-seeking commonplace \cite{greidanus2010online}. Most of the research conducted in this area focuses on health-related topics, including motivations, behaviors, and effectiveness of online help-seeking for different health concerns (such as physical health and mental health) and across a variety of demographics (such as adolescents, adults, and the elderly) \cite{nicholas2004help, kauer2014online, ybarra2006help, powell2011characteristics, miller2012online, de2014seeking, taylor2014evaluation}.

This paper extends research on online help-seeking to a crisis scenario. Many crises, such as flooding and pandemics, result in immediate losses and deaths of large populations and negatively impact their financial, physical, and psychological well-being in the long run \cite{gui2017managing, makwana2019disaster, yates1999help}. It is likely that affected people in such situations will require immediate assistance due to the physical danger or even the risk of death \cite{ullah2021rweetminer, nazer2016finding, yang2017harvey}. For example, they may request assistance with food, blood donations, and rescue individuals trapped \cite{ullah2021rweetminer}. Sometimes the size of disasters even overwhelms public emergency services, forcing individuals without adequate support to turn to interpersonal networks (such as friends or neighbors) or the Internet for assistance \cite{ekanayake2013we, reuter2012social, dreyfuss2015members}. However, help-seeking during an emergency is still a relatively new field of study. Many studies still focus on identifying urgent assistance requests and other emergency-related information on the Internet, as well as the development of systems to facilitate their transmission \cite{neubig2011safety, nazer2016finding, varga2013aid, kruspe2021detection,kaufhold2020mitigating,ohtsuka2017smartphone, sun2011rescueme, yang2017harvey}. In other studies, researchers have examined how individuals seek assistance after disasters (when the need is no longer urgent) \cite{yates1999help, labra2017men, urmson2016asking}. For instance, Urmson et al. \cite{urmson2016asking} examined 191 residents of New Zealand after an earthquake and found that people's comfort level with seeking assistance did not correlate with the level of support they received. 

However, current research does not adequately examine how people seek urgent assistance (especially online) during times of crisis. Based on a large dataset, we will provide empirical insight into this area by investigating help-seeking activities in Chinese social media during COVID-19.

\subsection{Crisis Communication}

During crises, people in need and emergency responders are connected by communication \cite{kapur2016effective}. While effective communication can reduce disasters' impact \cite{houston2015social}, traditional crisis communication channels such as radio and telephone-based systems are sometimes delayed or interrupted by massive help requests during a disaster \cite{jaeger2007community, shneiderman2007911}. For example, when Hurricane Harvey hit the Houston area in 2017, it was reported that 9-1-1 operators handled over 56,000 calls in 15 hours. Nevertheless, many calls were still unable to be answered \cite{yang2017harvey, li2019using}.

In recent years, new information and communication technologies (ICTs), such as social media, have opened up possibilities for better crisis response. ICTs enable two-way information flow between senders and receivers, with high capacity, dependability, and interactivity \cite{avvenuti2016framework, palen2007citizen, huang2010web, jaeger2007community}. In particular, social media allows users to consume content and create or influence it \cite{wright2009updated} and plays an increasingly important role in facilitating communication of self-coordination and assistance between affected groups and the public during times of crisis \cite{palen2009crisis, schultz2011medium, qu2009online, starbird2011voluntweeters, sutton2010twittering, reuter2018fifteen}. For example, Starbird et al. \cite{starbird2011voluntweeters} found that ``digital volunteers'' collaborated in self-organized ways to relay, filter, and verify information after the 2010 Haiti earthquake. In addition, emergency responders also use social media to share disaster-related information with the public to guide their actions when a disaster strikes \cite{reuter2017towards, bruns2012qldfloods, denef2013social}. For example, Bruns et al. \cite{bruns2012qldfloods} reported that the Queensland Police Service used Twitter to provide timely and useful information to the public after a flood, although these uses were largely ad hoc and unplanned.

However, current work does not provide a sufficient understanding of crisis communication via social media in terms of how disaster-affected groups request assistance and how crisis responders can help. This paper aims to fill this gap by exploring the effectiveness and limitations of social media, especially microblogging services, as tools to support such communication by using online help-seeking activities on Weibo during COVID-19 as an example.

\subsection{Microblogging for Help-Seeking During Crises}

During the past few years, social media has increasingly been used by individuals suffering from disasters as a tool to seek assistance. Microblogging sites such as Twitter, Facebook, WeChat, and Weibo are some of the most commonly used platforms \cite{nazer2016finding, rhodan2017please, li2019using, qu2011microblogging}. 

Microblogging enables users to share their status via short posts sent by instant messaging, mobile phones, email, or the Internet \cite{java2007we}. Microblogging platforms, \eg Twitter and Weibo, are increasingly considered as a novel emergent communication method due to its widespread usage and rapid communication capabilities \cite{vieweg2010microblogging}.
A number of microblogging hashtags were employed by people to request help during various disasters, and by volunteers and humanitarian organizations to provide assistance. For example, such hashtags used on Twitter include \#qldfloods (2011 Queensland Flood) \cite{bruns2012qldfloods}, \#ChennaiRainsHelp (2015 Chennai Rain), and \#PorteOuverte (2015 Paris Attacks) \cite{nazer2016finding}. 
Also, the \#Rescue hashtag has also been used on Twitter Japan in the 2011 Great East Japan Earthquake, the 2017 Northern Kyushu Heavy Rain, and the 2018 Japan Floods \cite{nishikawa2018time, Hiroyuki_FUJISHIRO2018, nishikawa2019analysis}.
In fact, the hashtag developed a fairly complete set of rules for help requests. @TwitterLifeline (an official account) instructed users to include their needs, address and image when making a request using this hashtag, and to delete the post once the assistance is received \cite{nishikawa2018time}. 
Using data from this channel, researchers have gained a preliminary understanding of the content and timing trends in emergency requests \cite{nishikawa2018time, nishikawa2019analysis}.
During the recent COVID-19 outbreak, an ad-hoc community (\ie \ST) emerged on Weibo to assist patients seeking medical help, which has in turn received much scholarly attention \cite{zhao2020online, huang2020mining, chen2021exploring, luo2020triggers}. Some studies have examined the demographic of patients in this community \cite{zhao2020online}, their geographic distribution \cite{huang2020mining, zhao2020online}, symptom characteristics \cite{huang2020mining}, and the commenting behavior of volunteers who provide support \cite{chen2021exploring}. One study has used this community to examine how social media affects government coordination \cite{zhao2020online}. Another study has analyzed 727 help-seeking posts on Weibo to determine the relationship between their content characteristics (e.g., emotion) and the number of reposts they received \cite{luo2020triggers}.  

The studies mentioned above, however, used limited data sources, where searches were conducted using only a single hashtag or a few keywords, thus restricting the scope of the study. There could be many requests that go unnoticed, such as those using different hashtags \cite{acar2011twitter}. By using a reliable classifier to examine over 100 million posts, our study contributes to a more comprehensive understanding of how disaster-affected populations seek help online.

\section{Background} \label{sec:bg}
\subsection{Crisis Event: Outbreak of COVID-19}

\begin{figure*}[h]
\includegraphics[width=\textwidth]{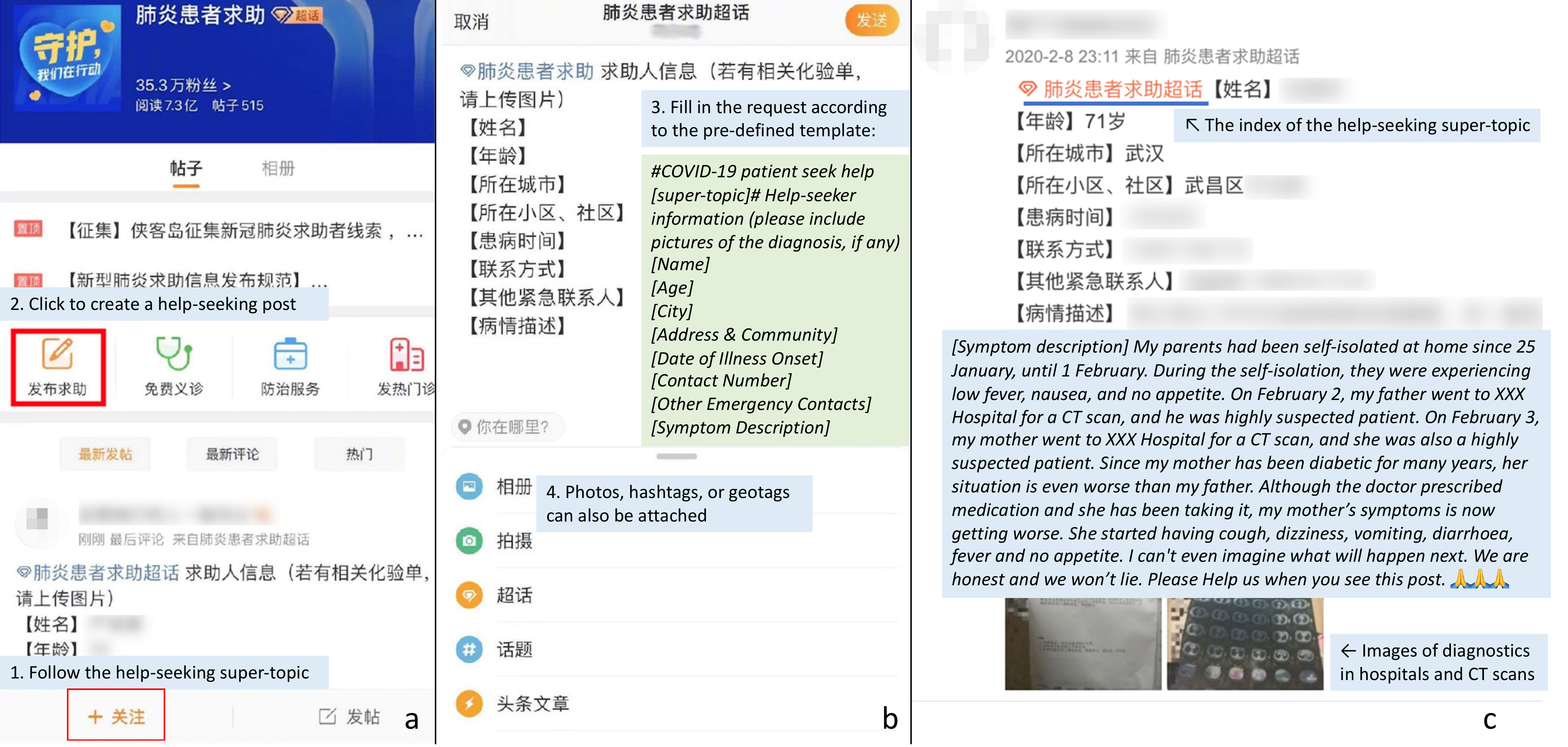}
\centering
\caption[Caption]{
Steps to use the help-seeking super-topic (Figure a and b) and an example request on this super-topic (Figure c). 
The figures a and b are taken from an official tutorial \cite{WCSO2020way} and the information shown in c was collected %
and being coalesced from multiple users to ensure privacy.
} %

\label{fig:bg-supertopic}
\Description{This figure is fully described in the text.}
\end{figure*}

In December 2019, a novel coronavirus (COVID-19) was firstly reported in Wuhan \cite{zhu2020novel}, the capital of Hubei Province, China, and quickly spread globally. 
The outbreak caused a conflict in medical resources and demand in Hubei, which led many people to struggle to gain healthcare. By April 8, 2020, there were 81,865 confirmed cases on mainland China, of which more than 80\% came from Hubei province alone \cite{NHCommission2020}. Healthcare in Hubei %
was severely overburdened as the number of infected individuals %
had risen dramatically \cite{khan2020novel}. 
On the other hand, policy factors also made it difficult for affected people to seek medical treatment on their own. In response to the outbreak, the Chinese government had placed Wuhan under lockdown since January 23. 
During the lockdown period, Wuhan's public transport and departures were suspended \cite{WuhanPCH1}, %
subsequent stricter policies prohibited vehicle usage in the city \cite{WuhanPCH9}, and citizens were not allowed to leave their homes without permission \cite{qian2021did}.  
Patients (and suspected patients with fever) must contact their communities (i.e., shequ) to arrange for treatment in a hospital or to be kept in home isolation for observation \cite{WuhanPCH10, WuhanPCH7}
.
Such ``grid-style social management'' and travel ban policies might have limit affected patients' access to timely medical care \cite{lin2020analysis, chen2020fangcang}, especially when inefficient management was present in some communities \cite{qian2021did}. As a result, many patients (and suspected patients) %
turned to social media to call for assistance.

\subsection{Research Site: Sina Weibo}

Sina Weibo is the most popular microblogging platform in China. It is reported that around 510 million people use it each month in 2020 \cite{WeiboUDR}. According to Sina's report %
\cite{WeiboST}, %
there were %
more than 200 million people following
information about the COVID-19 via Weibo every day during the pandemic, with users in Hubei growing by 34\% and posting 40 million posts. 
Weibo offers a variety of instant sharing mechanisms similar to those of Twitter. For example, users can share posts that include text, images, and videos, and utilize hashtags to categorize a post so that it will be more easily found in searches. It is possible for a hashtag to become trending if, for example, a large number of people use it within a short period of time. Also, users can follow others in order to view their followees' postings on their own timeline.

\ywj{Some features of Weibo are designed differently from those of Twitter. Particularly, Weibo has a kind of subreddit-like community called super-topic (will be described in Section \ref{sec:bg-supertopic}).} Also, the length of a regular post on Weibo is limited to 2,000 characters\footnote{When a post exceeds 140 characters, the excess will be collapsed and users will need to click a link to view the full content.}, which is much higher than Twitter's limit of 280 characters. This allows users to present a large amount of information in just one post. 
With these features of Weibo, it offers distinct advantages over other microblogging sites regarding help-seeking in times of crisis. For example, a super-topic community can set posting permissions to exclude irrelevant information and facilitate efficient centralized management of information related to help-seeking. Additionally, the help-seeking super-topic offers a template with very detailed requirements for users to fill out (will be described in Section \ref{sec:bg-template}). In contrast to sending a tweet, Weibo users may not need to worry about exceeding the word limit and can provide useful information for responders as much as possible. By using Weibo as a research site for research on crisis informatics, this paper will provide new insights into the existing literature, which was mainly focused on Twitter.

In this section, we will first introduce the super-topic function of Weibo that has been prominently used by people seeking help during crises, as well as the template used in super-topic posts.

\subsubsection{Super-Topic Community on Weibo} \label{sec:bg-supertopic}

A super-topic extends the functionality of  a hashtag by creating a long-lasting community for it which operates like a subreddit.
There will be a dedicated page for the community to display posts from its members, as shown in Figure \ref{fig:bg-supertopic}.a. Users must subscribe to the super-topic to become members and have permission to post there. As presented in Figure \ref{fig:bg-supertopic}.c, each super-topic post will carry a hashtag-like index, searchable through Weibo's search function. The posts are visible both inside and outside the community, but members of the super-topic will receive them directly on their personal timelines.
There can be an administrator for a super-topic, who is allowed to audit, filter, and pin posts, as well as edit the introduction %
and establish posting rules%
. Active followers can apply to become administrators.
Popular super-topic communities on Weibo include celebrities, TV shows, hobbies, etc. 

On January 29, 2020, Weibo launched the super-topic ``\ST'' (\#\begin{CJK}{UTF8}{gbsn}肺炎患者求助[超话]\end{CJK}\#), dedicated to COVID-19 patients seeking help.
\ywj{
This community quickly gained the attention of online volunteers. Specifically, a team of about 1,000 volunteers established an official partnership with Weibo to collect help requests from the platform (mainly monitoring the super-topic \cite{LHJY2020}) on a 24/7 basis \cite{WCA2020WAO}. During this response process, as illustrated in Figure \ref{fig:bg-info_flow}, the volunteer team communicated proactively with the help seekers (by contacting the telephone number provided on their posts), verified their situation, and referred cases to local non-governmental organizations (NGOs) and authorities who could arrange on-site assistance, such as the Hubei Provincial Internet Information Office \cite{BJQN2020}. 
}
\subsubsection{The Template of the Help-seeking Super-topic}
\label{sec:bg-template}

In the help-seeking super-topic, each post had to follow a structured \textit{template}, as shown in the green box in Figure \ref{fig:bg-supertopic}.b. 
When a user subscribes to this community and initiates a post, a predefined piece of content will be automatically embedded into the post. The template begins with a searchable index (i.e., \ST) and a suggestion for the user to upload images of their test reports. Following this are eight items enclosed in brackets that ask the user to provide: the \textit{name} and \textit{age} of the patient, the \textit{city} in which they lived, their complete \textit{address}, \textit{date of illness}, their \textit{contact number}, \textit{other emergency contacts}, and a description of their \textit{symptoms}. The user can enter their information after each bracketed item. An example of help requests is shown in Figure \ref{fig:bg-supertopic}.c. As the template is text based, it is technically possible for users to modify it or skip certain sections.
However, a director of Weibo indicated that posts lacking the necessary information required by the template would be deleted, ``\textit{The messages here [within the super-topic] will be exported to relevant departments, so the community address and phone number are required fields. Ambulance personnel is unlikely to [have time to] visit the internet to send you a private message. All posts without such information will be removed to conserve resources.}'' \cite{LQZJ2020} 

\section{Data Collection}

In this section, we show how we collect and integrate posts related to COVID-19 from Weibo, and then train a classifier to detect help requests in them.

\begin{table}[h]
\centering
\rowcolors{2}{white}{gray!15}
\resizebox{\columnwidth}{!}{%
\begin{tabular}{llll}
\toprule
Dataset     & Collection Time & \# Posts    \\ \midrule
Ours        & January 1 - May 18, 2020  & 136,803,626             \\
Weibo-CoV    & April 1 - May 4, 2020   & 40,893,832             \\
Wuhancrisis & February 12 - February 16, 2020 & 1,214                      \\ \bottomrule
\end{tabular}}
\caption{Summary statistics for Weibo data. 
}
\label{tab:dc-dataset}
\end{table}

In order to identify help requests related to COVID-19 patients on Weibo, we first collected a set of Weibo posts related to COVID-19, and then trained a machine learning algorithm to recognize requests from the dataset.
Our study is based on data taken from three sources, covering 177 million posts related to the pandemic between January 1 and May 18, 2020: 1) Our main dataset was provided by a commercial organization\footnote{\url{http://yuqing.gsdata.cn}}(no relationship with the authors). The data was gathered by tracking real-time posts on Weibo and then filtered by epidemic-related keywords (such as ``pneumonia'' and ``Wuhan''). This dataset also includes publicly available meta-data of accounts, such as gender and follower number. 2) Another dataset we used is Weibo-COV V1 \cite{hu-etal-2020-weibo}, an open-source dataset obtained from crawling 250 million active users' timelines and filtering the results based on keywords. 
3) We also used data from a website, Wuhancrisis\footnote{\url{https://www.wuhancrisis.com/}}, that contains help requests obtained from daily crawling of the super-topic homepage. 
The specific data collection time and the number of posts for each sub-dataset are shown in Table \ref{tab:dc-dataset}.

\subsection{Help Request Identification}

\subsubsection{Data Annotation}

Despite Weibo creating the super-topic at the end of January, we %
noticed some help requests have been scattered before the super-topic was created or outside the super-topic community. Such requests might not contain a constant hashtag or might be hard to filter out by fixed keywords. In order to recognize as many requests as possible on the platform, we built a BERT-based classifier. 
The classifier's input is the text of a post, and the output is a binary value indicating whether or not the post is a help request related to COVID-19 patients. We trained the classifier using a small subset of the labeled training data, then apply it to the entire dataset.

The authors first developed a codebook for identifying help requests using 200 randomly selected posts from the Wuhancrisis dataset (i.e., posts in the super-topic).
1) \textit{Requests related to COVID-19 patients}, including suspected as well as confirmed patients. 
In light of the definitions in past work \cite{ullah2021rweetminer, zhao2020online}, a help request must entail a request for some resource (e.g., medical resources or information).
Furthermore, we followed the ``Diagnosis and treatment protocol for novel coronavirus pneumonia'' \cite{national2020diagnosis} to determine whether a request was from a COVID-19 patient. The protocol describes the symptoms associated with COVID-19 (e.g., fever, diarrhea, and ground glass shadows in the lungs) used to diagnose during the outbreak. 
Whenever a request mentioned one of these symptoms, we categorized it under this code. Considering people's limited knowledge of the virus in the early stages of the crisis, we also classified those requests that did not provide symptoms but directly stated that they were infected or had a positive nucleic acid test result as this code.
2) \textit{Other posts}, including those not fitting into the first category, such as requests from non-COVID-19 patients, posts of sympathy, solidarity, and opinions from the public, and other posts related to the outbreak.
As help requests represent a very small percentage of the whole dataset, we did not sample random posts from the entire dataset for annotation. Instead, two authors randomly selected and coded 1K posts from the Wuhancrisis dataset, achieving a Cohen kappa of 0.826, and resolved disagreements after discussion. This obtained 783 positive examples (i.e., help requests) and 217 negative examples. Afterward, each author reviewed another 1K posts from the entire dataset, which were selected at random without overlap. The 2K posts were all found to be negative. The annotations resulted in 3K examples, of which 783 are positive, and 2,217 are negative. 

\subsubsection{Data Augmentation}
Since all positive examples follow the template of the help-seeking super-topic (whereas the negative examples do not), the above annotation may cause a classifier to rely exclusively on the template for determining help requests and making biased decisions. To alleviate this issue, we used data augmentation to generate a number of positive examples that do not conform to the template format. In particular, we removed all brackets (along with titles within brackets), names, ages, and phone numbers from the posts. Taking out these details will not change a positive example into a negative one, but modifying other information, such as symptoms, may do so. In addition, we utilized a medical translation API\footnote{\url{https://medtrans.damo.alibaba.com/medtrans.htm}} to translate these (unaugmented) posts into English and then back into Chinese to increase the diversity of positive results. This step can further enhance the generalizability of models.

In order to apply the two-step augmentation process, we divided the annotated corpus into training, validation, and test sets and augmented only the positive data from the training set. Ultimately, we obtained a total of 3,682 training, 300 validation, and 300 test data points.

\subsubsection{Model Performance and Result}

\begin{table}[]
\centering
\rowcolors{2}{white}{gray!15}
\resizebox{\columnwidth}{!}{%
\begin{tabular}{lllll}
\toprule
Model               & Accuracy & Recall & Precision & F1    \\ \midrule
BERT                & 0.960     & 0.967  & 0.953     & 0.959 \\
BERT+Augmented Data & \textbf{0.980}     & \textbf{0.984} & \textbf{0.975}     & \textbf{0.980}  \\ \bottomrule
\end{tabular}
}
\caption{Models' average performance in identifying help requests associated with COVID-19 patients}
\label{tab:dc-performs}
\end{table}

We adopted Bidirectional Encoder Representations from Transformers (BERT) \cite{devlin-etal-2019-bert} as a classifier. 
We adopted the Chinese BERT implemented by huggingface\footnote{\url{https://huggingface.co/bert-base-chinese}}. Using the test set, we compared BERT performance before and after augmentation. In Table \ref{tab:dc-performs}, it can be seen that the augmented model gains improvements in all metrics. F1 scores of this model on the positive and negative examples are 0.983 and 0.975, respectively.
With this model, we were able to locate 105,892 posts containing help requests in the 100 million posts on Weibo. We kept all the original posts among them. With the help of BERT, we discarded reposts in which the quoting content was a help request but the reposting comment was not, since they were generally about supporting from the audience. Posts from different datasets were de-duplicated based on their ID. 

Finally, we retrieved \totalCount help requests related to COVID-19 patients from 6,158 distinguished users. 
\section{Microblogging functions used for help requests (RQ1)} \label{sec:rq1}

\ywj{
In order to learn what microblogging functions of Weibo were used for help-seeking and how effective they are, we analyzed the function usage of 8K help requests and examined their affordances and limitations both qualitatively and quantitatively.
}

\paragraph{Function Identification} \label{rq1:identifyFunctions}

Two researchers reviewed our dataset, identifying which functions COVID-19 patients or their caregivers employed to increase their chances of receiving assistance on Weibo. Using an iterative approach similar to that used by Krafft et al. \cite{krafft2017centralized}, they identified functions used in the help requests. They analyzed the content and metadata (e.g., if a {post} was original) of 100 randomly sampled requests in each round, coding the microblogging function(s) involved in these posts. They held discussions to iteratively refine the proposed codes; for instance, similar or duplicated features were combined into one mechanism, whereas some other cases may be divided into more specific mechanisms. In the next round, they used the codebook derived from the previous round to explain or discover new features based on 100 more {posts}. They repeated the process for four rounds until no more new features were discovered. 
Eventually, a total of 11 types of microblogging functions were recognized as described in Table \ref{tab:rq1-functions}. 
Note that three functions (\ie \textit{geotag, article, private message}) are not analyzed in the subsequent quantitative analysis of RQ2 and RQ3 due to the lack of complete data about their usage in our dataset.

After that, we labeled the use of different functions in the full 8K requests. We wrote programs to automatically detect some of the function usages. For example, {posts} are determined to be using the \textit{help-seeking super-topic} if they match the format of the super-topic's template (as shown in Figure \ref{fig:bg-supertopic}.c). An original post (indicated by its metadata) will be classified as using the ordinary \textit{posting} function. If the quoting content of a {repost} is not classified as a help request by the previous BERT classifier, but the comment added by the current {reposting} user is, %
this post will be considered as using the \textit{quoting} function. 
Examples of requests via quoting can be found in Section \ref{Quoting}. %
We checked these posts manually for the possible usage of auxiliary  
as well. If and only if a post uses a hashtag after this hashtag appeared in the trending list, %
the post will be labeled as using a \textit{trending topic}. %
By checking if an account mentioned ($@$) in a help request actually exists or if this account has more than 1,000 followers, we determine whether a post \textit{misuses mentions} or \textit{mentions influencers} respectively.
Finally, we manually inspected all hashtags and super-topics to compile a list of the ones that are misused as help-seeking topics. Using this list, we can automatically detect posts that are \textit{misusing the help-seeking super-topic} or asking for help in \textit{unrelated super-topics}.

\begin{table*}[hbt!]
\centering
\rowcolors{2}{white}{gray!15}
\resizebox{0.8\textwidth}{!}{%
\begin{tabular}{ll}
\toprule
Function (N)                    & Description                                                                              \\ \midrule
\textbf{Standalone} & \\
               \ \ Help-seeking super-topic (3652)         & 
               Post in the \ST \\
               \ \ Unrelated super-topic (27)                 &  \begin{tabular}[c]{@{}l@{}}
               Post in super-topics that are not primarily used for helping \\COVID-19 patients 
               \end{tabular}\\
               \ \ {Posting} (4307)             & Create an original post in personal timeline                                                      \\
               \ \ Quoting (227)             & Quote ({repost}) certain news {posts} along with comments about a request                                           \\
               \ \ Article (5)                 & Use Weibo's headline to publish an article                                                 \\
               \ \ Private message (/)              & Send private messages to influencers and seek help                                 \\ \midrule
\textbf{Auxiliary} & \\
               \ \ Trending hashtag (784)                       &  Use hashtags which are or have appeared in trending topics                                                                                                 \\
               \ \ Mention (2065)           & Mention (@) an influencer (follower count > 1K) to get their attention                               \\
               \ \ Geotag (/)                         & Tag geographical location to the {post}                                                           \\ \midrule
\end{tabular}}
\caption{The 9 types of microblogging functions used by help seekers to post requests, categorized as standalone and auxiliary classes.\\
}
\label{tab:rq1-functions}
\end{table*}

\subsection{Function and Its Misuse}

Table 3 summarizes six primary features, or \textit{standalone functions}, available for creating help requests on Weibo. Depending on the function, requests will be presented in different formats to different audiences. For example, one can publish a post or an article that contains help-seeking messages for their followers or send a private message to specific help providers. Also listed in the table are three \textit{auxiliary functions} that do not directly contribute to the creation of content, but are able to provide a broader audience reach when combined with standalone functions. For instance, a post that includes trending hashtags is also likely to be seen by users visiting the trending pages, in comparison with a plain post.

In analyzing the practices of users using Weibo services, we noted that some incorrectly use (\textit{misused}) the help-seeking super-topics and @ mentions. When misused, the functions will not work as expected. For example, posts that misuse the super-topic feature will not be published within communities, but only on users' timelines, causing posts to be less visible. By ``misuse,'' we do not imply that users intentionally use the functions improperly. In fact, our following analysis indicates that such behaviors are likely a consequence of users experiencing technical difficulties when attempting to utilize functions.
Below, we will discuss each function in detail.

\subsubsection{Super-topic}
\label{super-topic}
As described in Section \ref{sec:bg-supertopic}, super-topics are topic-based communities on Weibo, which have their own subscribers and administrators.
According to our data, help requests related to COVID-19 patients have been found both in dedicated help-seeking communities (i.e., \ST) as well as in other communities that are not specifically used for help-seeking. 

\paragraph{Help-Seeking Super-topic: Dedicated Management and Great Attention}
This community was established exclusively for COVID patients seeking assistance, as it stipulated that ``\textit{Users who do not seek help are not allowed to post in this community; violators will be banned for one week}'' \cite{WADMIN2020}. 
\ywj{
As described in Section \ref{sec:bg-supertopic}, Weibo cooperated with volunteer teams to assist with community management by handling and referring help-seeking messages to local authorities. But we do not know whether these volunteers were granted special administrative or moderation rights on the platform. 
}
We found that almost half (43.50\%) of the \totalCount help requests we identified were submitted through this channel. Nevertheless, the ratio may still be underestimated. According to a Southern Weekend report \cite{FENG2020SW}, strict management of this super-topic in the first few days after its launch (February 3 to February 4) may have prevented some posts from becoming public:
``\textit{after Weibo's vetting, the final help-seeking posts visible to the public [in this super-topic] are 20 percent of the total.}''
Weibo administrators clarified on February 4 that ``\textit{posting in this super-topic does not require authentication, and {posts} can be sent out as long as they contain all the information needed. Please do not publish duplicate {posts}, as the relevant departments will verify and handle them}'' \cite{STC2020}.
If posts are made public, they can be viewed by users who subscribe to the super-topic as well as those who search the super-topic. 
The super-topic has drawn considerable attention from the public. As of August 29, 2021, the posts on the super-topic have been viewed 5.1 billion times in total.

\label{sec:rq1-misusedST}

\begin{figure*}[t]
\includegraphics[width=0.8\textwidth]{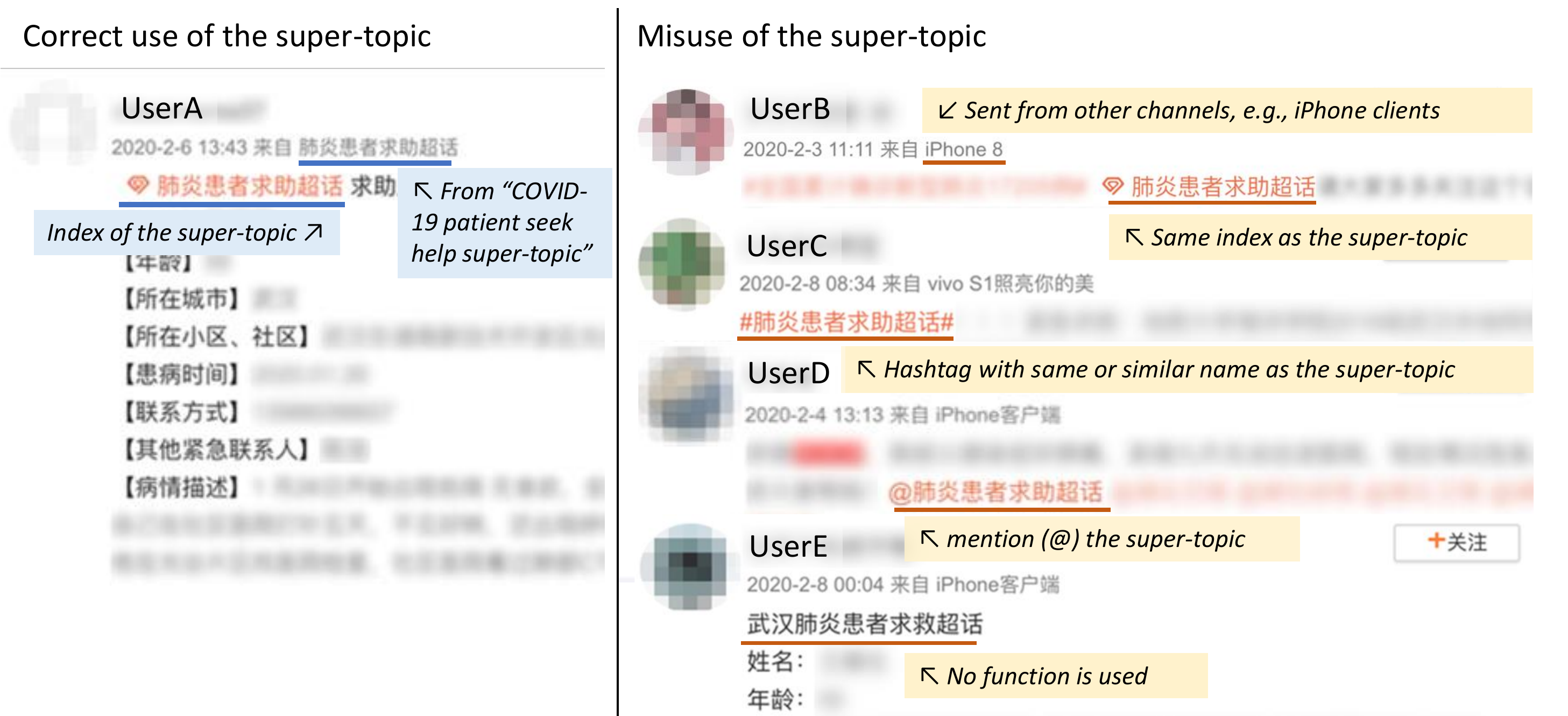}
\centering
\caption{
Examples of the correct use of the help-seeking super-topic and the four types of misuses we found.
}
\label{fig:rq1-misusedst}
\Description{This figure is fully described in the text.}
\end{figure*}

While users may want to post their help requests in this specialized super-topic, some of them seem to apply the super-topic function incorrectly. We found 177 such requests that misused the super-topic hashtag and thus did not appear in the community. Some of the typical errors are listed below.
Some users did not open the super-topic home page and subscribe it (without carrying out the first step in Figure \ref{fig:bg-supertopic}), but instead, they manually typed the super-topic index ``\ST'' into the {post} content, as if they were using an ordinary hashtag. As a result, the super-topic index (similar to the form of a hashtag) appeared within the post but did not actually let the {post} be posted in the community.
\ywj{
Figure \ref{fig:bg-supertopic} (right) shows that even if UserB (a misuser) applies the same super-topic index as UserA (a correct user), UserB's metadata indicates that the message was actually posted on the timeline (in this case, the Weibo interface would display users' device information by default, e.g., ``iPhone clients'').
}
Many other users 
attached the wrong hashtags, such as ``\#COVID-19 patient seek help super-topic\#'' (missing ``[]'', as in UserC's post) and ``\#COVID-19 patient seek help\#'' (missing ``[super-topic]''). We found 13 %
misused hashtags similar to the help-seeking super-topic's index.
Even worse, some users who were perhaps unfamiliar with hashtag usage used the mention (\ie @) function instead to reference the topic's name, or just typed the topic name out, as illustrated in the Figure \ref{fig:rq1-misusedst} for UserD and UserE. 
Such incorrectly used hashtags can greatly limit the attention from and access of potential audience for a post \cite{alsini2021hashtag}; for example,
non-followers
will not see these {posts} on their timeline. As of August 29, 2021, the aforementioned two mistaken hashtags received between 60 million and 300 million reads, which is far fewer than the 5.1 billion reads on the correct super-topic. 

\paragraph{Unrelated %
Super-topic}
We found a small number of help-seeking posts (N=27) 
posted to super-topic communities that were not specifically set up for help-seeking, including those associated with a geographic region (e.g. Wuhan), medical facility (e.g., Huoshenshan Hospital built for the epidemic), celebrity, and other Internet influencer. For example,
``
\textit{\#Huoshenshan Hospital [super-topic]\#
My name is XXX. 
... So far, I have not been arranged to seek medical treatment and hospitalization. ... I hope kind people can forward this to the government to arrange for my hospitalization. 
Thank you very much! 
Phone number XXX}.
''
We did not observe any specific changes in the content of these requests with respect to the type of community they appeared in. Several such requests were simply copied and pasted in the different communities. 

\subsubsection{Posting: More Freedom but Less Attention Received}
\label{tweeting}
Creating an original post (like tweet in Twitter) is a fundamental feature of Weibo. Users can create a {post} from their individual dashboard, and the {post} will be displayed in their timeline and seen by their followers. Surprisingly, we discovered that the largest number of help requests (N=4307, 51.30\%) were generated in this manner, even more than those created in the help-seeking super-topic.
By {posting}, users are free to choose the format and information they would like to disclose (or not). For example, people might not want to reveal their real names.
``\textit{My friend's family in Wuhan are infected, but without nucleic acid tests, they cannot be diagnosed and receive hospitalization. The situation is urgent. Help!!! }''
``\textit{\#COVID-19 confirmed more cases than SARS\# Please help, my classmate's father continued to have fever for 6 days. ... He went to three hospitals every day without being able to arrange hospitalization...}''
However, the disadvantage of this approach is apparent as well. {Posting} tends to attract a smaller body of audience than the super-topic, because the majority of viewers for a {post} would be the author's followers. It is also not guaranteed that all help-wanted posts in this form would receive the same amount of exposure. Although Weibo's keyword search engine may capture these {posts}, they are likely to submerge in the large volume of other posts with similar context circulated at every moment of the crisis.

\subsubsection{Quoting}
\label{Quoting}
Similar to Twitter's quote feature, Weibo allows users to {repost} a post and add their own comments on top of it. We observed that many help requests (N=227) were delivered by quoting trending news. Interestingly, these news are often related to immediate needs of COVID-19 patients. For example: \\
``[COMMENT1] \textit{Save us! My colleague's father has been diagnosed as a confirmed case for five days, but we can not find a sickbed for him.}
[NEWS1] \textit{\#Wuhan plans to build three Fangcang shelter hospitals\# ... The `Fangcang shelter hospital' will have 1,000 beds at the Wuhan International Convention and Exhibition Center.}''\\
``[COMMENT2] \textit{Please help us! Someone in my family has been diagnosed as a suspected case, but we can only stay at home due to the lack of sickbeds.}
[NEWS2] \textit{\#Wuhan's special inspection of infected patients\# Wuhan set up a special inspection team to inspect the condition of infected patients. Patients `should be accepted as much as possible'...}''\\
``[COMMENT3] \textit{I have not received any assistance so far! No driver is willing to take me to the hospital.}
[NEWS3] \textit{Good news! On the afternoon of the 2nd, Jinyintan Hospital discharged 37 COVID-19 patients...}''

\begin{table*}[]
\centering
\resizebox{\textwidth}{!}{%
\rowcolors{2}{white}{gray!15}
\begin{tabular}{lll}
\toprule
Trending Topics (N=1,376)                                          & Mentioned Users (N=10,790)                   & Mentioned Non-existent Users (N=1,749)                 \\ \midrule
\#Nationwide Confirmed Cases of COVID-19\# (15.48\%)      & @People’s Daily (9.48\%)               & @Wuhan (8.06\%)                                    \\
\#Be Strong Wuhan\# (9.23\%)                                   & @CCTV News (7.98\%)                    & @Wuhan Health Committee (7.95\%)                   \\
\#The Girl who Knocked Gong\# (8.65\%)                         & @People’s Daily Online (6.13\%)         & @Wuhan Jinyintan Hospital (7.66\%)                 \\
\#Wuhan\# (3.12\%)                                             & @The Central Hospital of Wuhan (5.08\%) & @Wuhan Government (6.29\%)                    \\
\#Latest Epidemic Map\# (2.33\%)                               & @Wuhan Evening News (3.90\%)            & @Wuhan Pneumonia Prevention Center (5.03\%)        \\
\#Wuhan Diary\# (2.33\%)                                       & @Safety Wuhan (3.75\%)                  & @Wuhan Hongshan District Government (5.03\%) \\
\#The Epidemic is still in the Spreading Stage\# (1.38\%) &
  @China Central Television Online (3.00\%) &
  @Provincial Government (2.97\%) \\
\begin{tabular}[c]{@{}l@{}}\#Those who Refuses to Cooperate with Quarantine  \\will Face Forcible Execution in Wuhan\# (1.16\%)\end{tabular} & \begin{tabular}[c]{@{}l@{}}@Xinhua News Agency Online (2.92\%) \end{tabular}  &
\begin{tabular}[c]{@{}l@{}}@Ezhou City Government (2.92\%) \end{tabular}\\
\begin{tabular}[c]{@{}l@{}}\#Vice Governor of Hubei Responded to Wuhan \\Citizens' Online Help-seeking\# (0.94\%)\end{tabular} &
 \begin{tabular}[c]{@{}l@{}} @Chinese government website (2.81\%) \end{tabular}&
  \begin{tabular}[c]{@{}l@{}}@COVID-19 Medical Team (2.17\%) \end{tabular}\\
\begin{tabular}[c]{@{}l@{}}\#More Confirmed Cases of COVID-19 than SARS\#\\ (0.87\%)\end{tabular} & @Wuhan Promotion (2.28\%)             & @Bai Yansong (1.14\%)                              \\ \bottomrule

\end{tabular}%
}
\caption{Top 10 trending hashtags, mentioned users, and mentioned non-existent users. N in each column represents the number of times a particular feature is used, not the number of requests that contain it.
\\
}
\label{tab:rq1-examples}
\end{table*}

\subsubsection{Hashtag} 
In our dataset, excluding the hashtags mistakenly used as the super-topic, a total of 809 posts included some hashtags, and almost all (N=784) used trending or previously trending hashtags (see examples in Table \ref{tab:rq1-examples}). This usage preference may reflect help seekers' desire to gain exposure for their requests. In addition, it is possible that trending topics were noted more frequently by users due to convenient access, since both Weibo and Twitter's interfaces display such topics readily on their dashboards \cite{wukich2013nonprofit}.
As shown in Table \ref{tab:rq1-examples} (left column), the popular hashtags being used do not seem to have a lot in common except being associated with the epidemic.

\begin{figure}[t]

\includegraphics[width=\columnwidth]{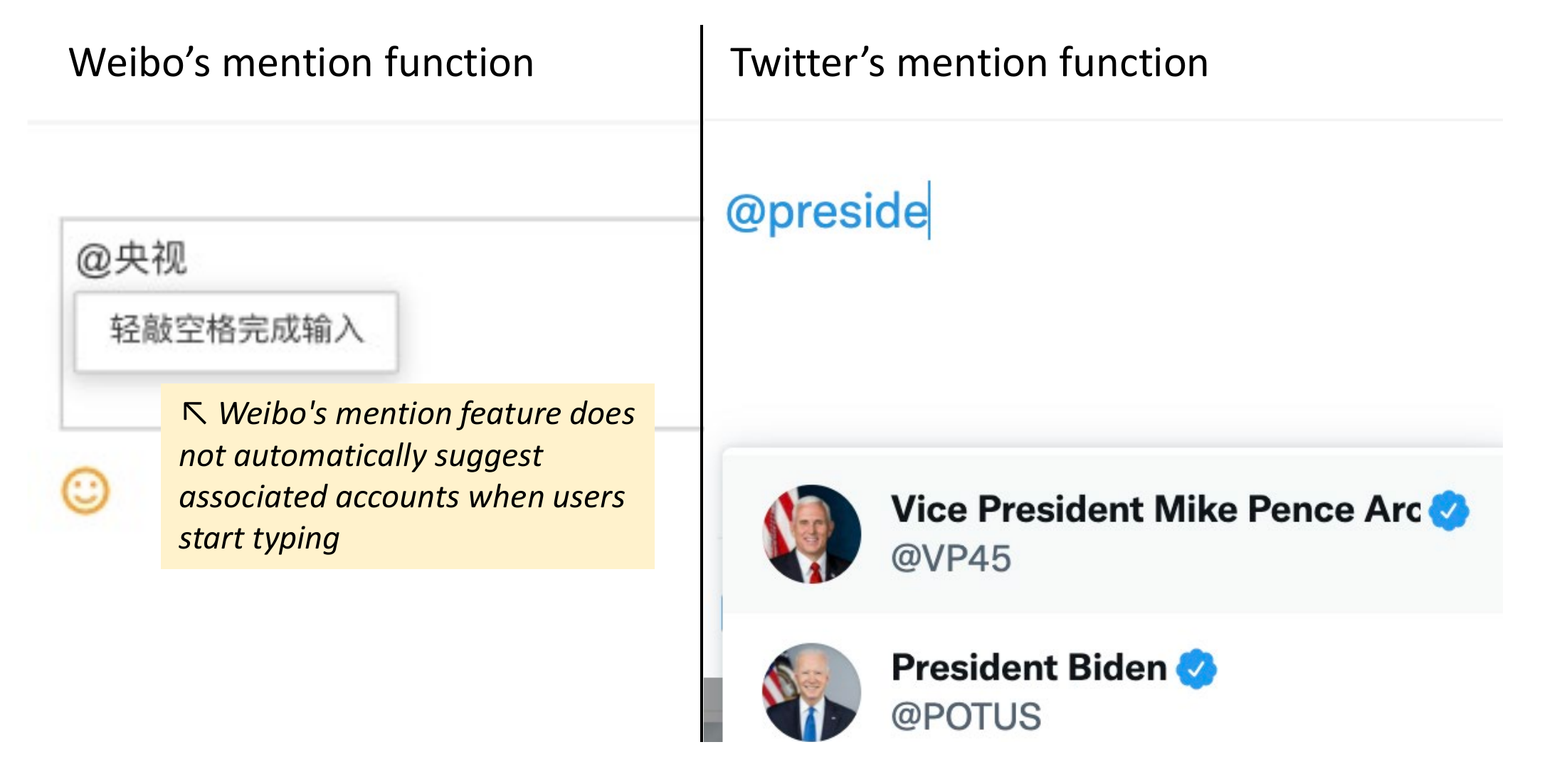}
\centering
\caption[Caption]{
Comparison of the mention function in Weibo and Twitter. 
}
\label{fig:rq1-mismention}
\Description{This figure is fully described in the text.}
\end{figure}

\subsubsection{Mention}
\label{mention}
In a {post}, the mention function can be used to engage with other users, attract attention, and increase visibility. Mention was the most common auxiliary feature used by help seekers. A total of 2,065 posts $@$ Weibo users with large follower bases,
demonstrating that people are eager to make their voices heard by these influencers. 
As shown in Table \ref{tab:rq1-examples} (middle column), national and local media in Hubei are referenced most frequently, along with hospitals and some opinion leaders. However, this function's usefulness is hard to evaluate because we cannot know whether those mentioned actually responded and whether they acted after becoming aware of it through the @ notifications or by other means. For example, Tong Zhiwei, a professor at East China University of Political Science and Law who has over 400K followers and actively helped {repost} requests on Weibo, said that he accessed such information primarily by email and private messages on Weibo \cite{LY2020TZW}.

Unexpectedly, we also noticed that 309 {posts} mentioned user accounts that do not exist. Examples are shown in Table \ref{tab:rq1-examples} (right column).
While this may indicate help seekers are eager to receive answers from professional aid organizations, hospitals, and local governments, their calls are less likely to be answered. This type of mistake might be caused by the limited experience of users and the lack of error prevention support by the Weibo platform. Unlike Twitter, Weibo will not prompt for the existence of associated accounts when users type their name after @, as shown in Figure \ref{fig:rq1-mismention}.

\subsubsection{Other functions}

While going back to the website to track some of the {posts}, we noticed several additional functions related to help-seeking that were not captured by our dataset.
\textit{Geotagging} allows users to share their location in {posts}. The requests  
flagged by a geotag are visible to
other users in the same city, thereby potentially increasing the reach. By showing the location of help seekers in the Hubei province through geotagging, it may also enhance the credibility of their requests.
Additionally, we found a few users asking for help via Weibo's headline, 
a feature that allows users to share \textit{articles} within {posts}, affording greater editorial freedom.
Also, we noticed that some influencers claimed to have received help requests in \textit{private messages}. Channels of this type are better for ensuring the privacy of help seekers.
These data were not collected in our dataset, so their analysis will not be included in the following quantitative analysis.

\subsection{Functions Differences}

In this subsection, we differentiate these microblogging functions in terms of their efficacy for requesting help in crisis situations and their limitations. 

\subsubsection{Efficacy of Microblogging Functions From a Information Diffusion Perspective} \label{sec:rq1-efficacy}

Considering that help requests under a crisis situation often need to be spread to get noticed and supported by potential help providers, we measured the effectiveness of using microblogging for help-seeking from the perspective of information diffusion.  Following the setting of Vieweg et al. \cite{vieweg2010microblogging}, we conducted a chi-square test to determine the relationship between the type of microblogging function employed to create requests and whether they were {reposted} (denoted by a binary variable).
For standalone functions, we only consider the two most used ones, \ie the help-seeking super-topic and {posting} (represented by a binary variable) and exclude quoting, as explained in Section  \ref{rq1:identifyFunctions}. 

Despite that our dataset contains metadata on {reposts} for the 100 million {posts} collected, it is possible that some posts were {reposted} after the dataset was constructed. Therefore, we checked the 8K requests again on Weibo in June 2021 to get the final {repost} counts of those that were not deleted (N=1,472). 
We combined the data from both sources to determine if a request was {reposted}.
We left out unrelated super-topics due to the small sample size (N=27) in this category. In addition, since posts that misused the super-topic function are essentially equivalent to regular {posts} as they appear in a user's timeline instead of in the super-topic community, we thus merged them into the {posting} category in this analysis. Similarly, the misuse of mention is categorized as not using mention in the subsequent reporting. 

\begin{figure}[t]
\includegraphics[]{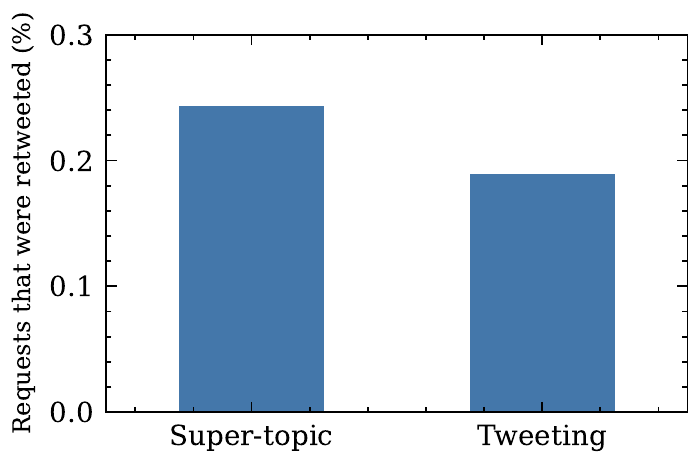}
\centering
\caption[Caption]{
The percentage of requests using the help-seeking super-topic and {posting} function that were {reposted}.
}
\label{fig:rq1-retweetability}
\Description{This figure is fully described in the text.}
\end{figure}

\paragraph{Dissemination (Reposting)}

The chi-square test found a significant association between the use of standalone functions and whether requests were {reposted}, $\chi^2(1)=35.68, p<0.001$. 
As shown in the 
Figure \ref{fig:rq1-retweetability}, 24.37\% of the requests that were created through the super-topic (N=3,652) were forwarded, whereas 18.88\% of requests created through regular {posting} (N=4,484) were forwarded. This result is consistent with what we discussed previously in Section \ref{super-topic} and \ref{tweeting}, %
 that posts in super-topics are more likely to attract attention than general {posts} due to their bigger follower (subscriber) base.

\ywj{
For auxiliary functions, we found that the use of hashtags (compared to no hashtag use) did not significantly associate with getting reposted ($p>0.05$), but the use of mentions did ($\chi^2(1)=12.50, p<0.001$).
} 
In particular, among the posts without a mention (N=6,326), 22.92\% were {reposted}, compared to 19.18\% with a correct mention (N=2,065), which suggests the use of mention may have a negative impact on predicting whether a post will receive a {repost}. This finding is in line with that of Suh et al. \cite{suh2010want}, that mentioning tends to reduce diffusion as it targets specific rather than broad audiences \cite{van2015police}. 

To sum up, requests posted using the help-seeking super-topic were more likely to be {reposted} than requests posted via regular {posting}, and including auxiliary features in the post did not appear to increase the likelihood of {reposting}.

\subsubsection{Limitations of Using Weibo for Help Requests}
\label{limitations}

There are some advantages of using each of the above features of Weibo to request help, such as the possibility of posting information to the public space in a timely manner (except for the super-topic that requires vetting) to be seen by humanitarian organizations or volunteers as reported by previous work \cite{chen2021exploring}. %
However, we have also identified a couple of limitations emerged with using Weibo for help-seeking in crisis situations.

\paragraph{Request Search Capability}
Help requests are only likely to be answered if the public and rescue organizations are aware of and able to locate them. Studies indicate that the search functions of microblogging platforms are ineffective for crisis communication \cite{palen2018social, ullah2021rweetminer,kapoor2018advances}, and our findings concur with this observation. Our analysis indicated that an unexpectedly high number of requests made via the Weibo {posting} function was difficult to be directly retrieved through hashtags or keyword searches provided by Weibo.
\ywj{
For example, a previous study by Huang et al. \cite{huang2020mining} used the keyword ``COVID-19 patient seek help'' (\begin{CJK}{UTF8}{gbsn}肺炎患者求助\end{CJK}) to identify help-seeking posts via Weibo's search API. However, we found that only 3,256 of the 8K requests we detected contained this keyword, suggesting that this method may not capture more than half of the overall platform's requests. Similarly, Luo et al. \cite{luo2020triggers} used a combination of three sets of keywords to extract help-seeking messages on Weibo, which only covered 6,439 (64.79\%) requests of our data. 
Additionally, feeding more keywords to the search engine may dramatically increase the likelihood of retrieving irrelevant posts, and thus greatly increasing the cost of filtering out 
such unrelated posts.
}

In comparison, locating help-seeking posts would be much easier if they were posted to a dedicated community. Using the two search options described above, 70.10\% and 73.78\% of the requests in the help-seeking super-topic could be collected, respectively. In such communities, the signal-to-noise ratio is much higher
than that of ordinary hashtag communities.
This is also reflected in our annotation process: 78.30\% of the 1,000 sample super-topic posts were relevant information (i.e., help requests). 
Furthermore, Acar et al. \cite{acar2011twitter} suggest that platforms should set up official hashtags specifically for crisis-affected people. This is because hashtag, a feature supported by all microblogging sites, is perceived as inferior for crisis communication due to the proliferation of multiple hashtags with different lengths and their varying abilities to contextualize information \cite{potts2011tweeting, acar2011twitter}. 

\paragraph{Request-Response Tracking Capability}
From the standpoint of 
\ywj{help providers such as the volunteers on Weibo,
}
getting feedback from help seekers when responding to their requests allows them to evaluate if sufficient support has been delivered, 
\ywj{
thus preventing waste of public resources.
}
Currently in Weibo, tracking the progress of crisis responses largely depends on help seekers actively updating the situation on their timelines via publishing follow-up information or re-editing their original posts. In reality, majority seekers just delete their help-related posts and do not provide any feedback. Other microblogging platforms also face similar issues and tried to guides users to make request-response tracking easier for help providers. 
For example, @TwitterLifeline, an active official user in \#Rescue, encourages victims to report back to them once they are rescued and then take down the request tweet that has been met \cite{nishikawa2019analysis}. 
This approach may ease help tracking to some extent; for instance, deleted posts cannot be retweeted and thus alleviating the problem of urgent requests being buried under a large amount of expired requests as found by Acar et al. \cite{acar2011twitter}. Our observations on Weibo, however, indicate that this approach are sometimes ineffective %
because from time to time posts about the same case are published in various places by multiple volunteers under different functions. There are no connections among these posts and it is difficult to track and update them all. 
\ywj{
One example is that some Weibo users eagerly helped those unfamiliar with the super-topic feature republish these requests (by creating new threads) in the help-seeking super-topic. 
}
The problem with this is that even if the help seekers have received necessary medical attention and deleted the original posts on their timeline, \ywj{threads} about their cases circulating in other channels may still be kept and reposted, resulting in a waste of resources. In some cases, even the seekers are not entirely certain who have published their requests and where. 
Currently, Weibo is unable to keep track of the duplicate requests spread around, even though the posts are often republished with little modification, according to our observation.

\paragraph{Ease of Use especially in Urgent Settings}
In the process of asking for help, misuse of microblogging functions can lead to failure in reaching the most relevant audiences and consequently reduces the likelihood of receiving assistance, as we discussed in Section \ref{sec:rq1-misusedST}. For one thing, Weibo does not offer adequate error prevention and recovery mechanisms, in particular, for the use of super-topic and hashtags. In the current implement of the platform, users can create a hashtag (\#keyword\#) and a super-topic index (\#keyword [super-topic]\#) in a similar way. They are both clickable and will take users to a page that shows the associated posts. Such similarities in visual form and interaction can confuse users, as our findings indicate. Weibo's mention feature also can not prevent users from referencing a non-existent account. 
For another, some of the Weibo functions are not very straightforward for the lay public. The practice of super-topics on Weibo shows that even when an officially endorsed help-seeking channel exists, users may be unaware or unfamiliar with its use. For example, many people do not know that they need to subscribe to a super-topic community first to publish any content in it. 
Such functionality deficiencies and inadequate usage guidance can reduce the efficiency of help-seeking.

\paragraph{Privacy Protection}
During emergency situations, privacy protection does not always seem to be a top priority in order to ensure efficient communication \cite{BPF2018CUSM}. 
Even so, the risk of privacy disclosure cannot be ignored. In this COVID-19 crisis, some help seekers on Weibo were reported receiving fraudulent phone calls, such as those selling counterfeit drugs, because they posted their phone numbers online when asking for medical assistance \cite{STC2020}. Weibo has been less sensitive to protecting the privacy of help seekers in this crisis. The most notable example is that posting to a super-topic requires users to disclose some very sensitive information, such as their address and phone number. Once published, this information is publicly accessible to all Weibo users and may be archived by search engines even if the author later deletes it.

\ywj{
In summary, we found multiple factors that affect the quality of experience when using microblogging for help requests and responses, including search and tracking capabilities, ease of use, and privacy protection. 
}
We present design considerations concerning each of these aspects in the discussion section.

\section{Usage of microblogging functions (RQ2)}

In RQ2, we investigate how people use the two most prominent microblogging functions (i.e., the help-seeking super-topic and {posting}) for making requests. Our first step is to visualize trends in the usage of these functions by help seekers over time and highlight two important events that may influence those trends, namely the creation of the super-topic and its exposure on the trending list. 
We then apply controlled interrupted time series (CITS) analysis to explore the possible intervention effects of super-topic exposure on the changes in people's use of these two functions.  
\subsection{Function Usage Over Time} \label{sec:rq2-evolution}

\begin{figure*}[h]
\includegraphics[width=.8\linewidth]{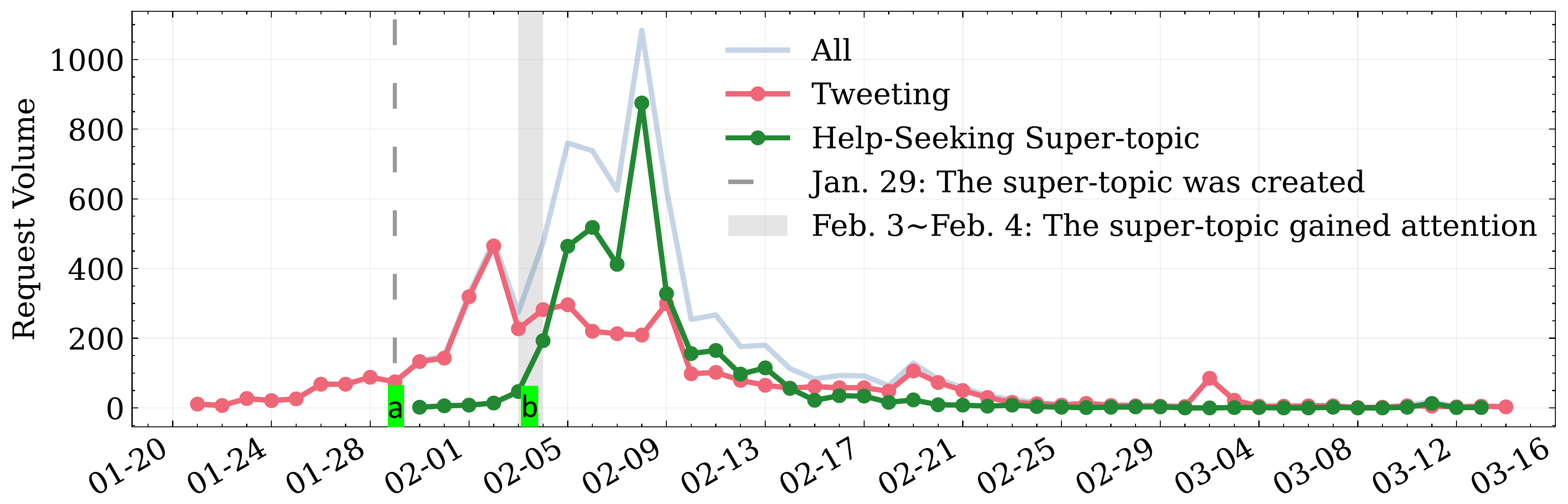}
\centering
\caption{
Number of help requests using different microblogging functions over time. 
}
\label{fig:RQ2-time}
\Description{This figure is fully described in the text.}
\end{figure*}

Figure \ref{fig:RQ2-time} illustrates the daily trend in the total number of help requests (light blue curve), 
as well as the trends in the number of requests via {posting} (red curve) and the help-seeking super-topic (green curve).
The earliest help requests appeared
on January 21, 2020, two days before the Lockdown of Wuhan city. It was not until the evening of January 29 that the help-seeking super-topic was created (Point a in Figure \ref{fig:RQ2-time}). The super-topic did not get much use until February 3, even though there had already been many help-seeking messages at the time (where the red lines always higher the green lines between points a and b). More than 70\% of all help requests were created between February 1 and February 10. After February 19, the daily number of requests declined to below 100. Help-seeking activities lasted 54 days and ended on March 15, 2020, when Weibo announced that the requests from the super-topic had been handled almost entirely and it would no longer report new requests to the authorities round-the-clock.
In general, 94.81\% of the requests were made via {posting} or the help-seeking super-topic. In particular, prior to February 5, 87.59\% of the 2,230 requests captured in our dataset were published via {posting}, but since then, 57.03\% of the 5,729 requests were created via super-topic.
Next, we introduce two important events in the life cycle of the help-seeking super-topic to analyze the change in Weibo function usage over time.

\begin{figure}[h]
\includegraphics[width=\columnwidth]{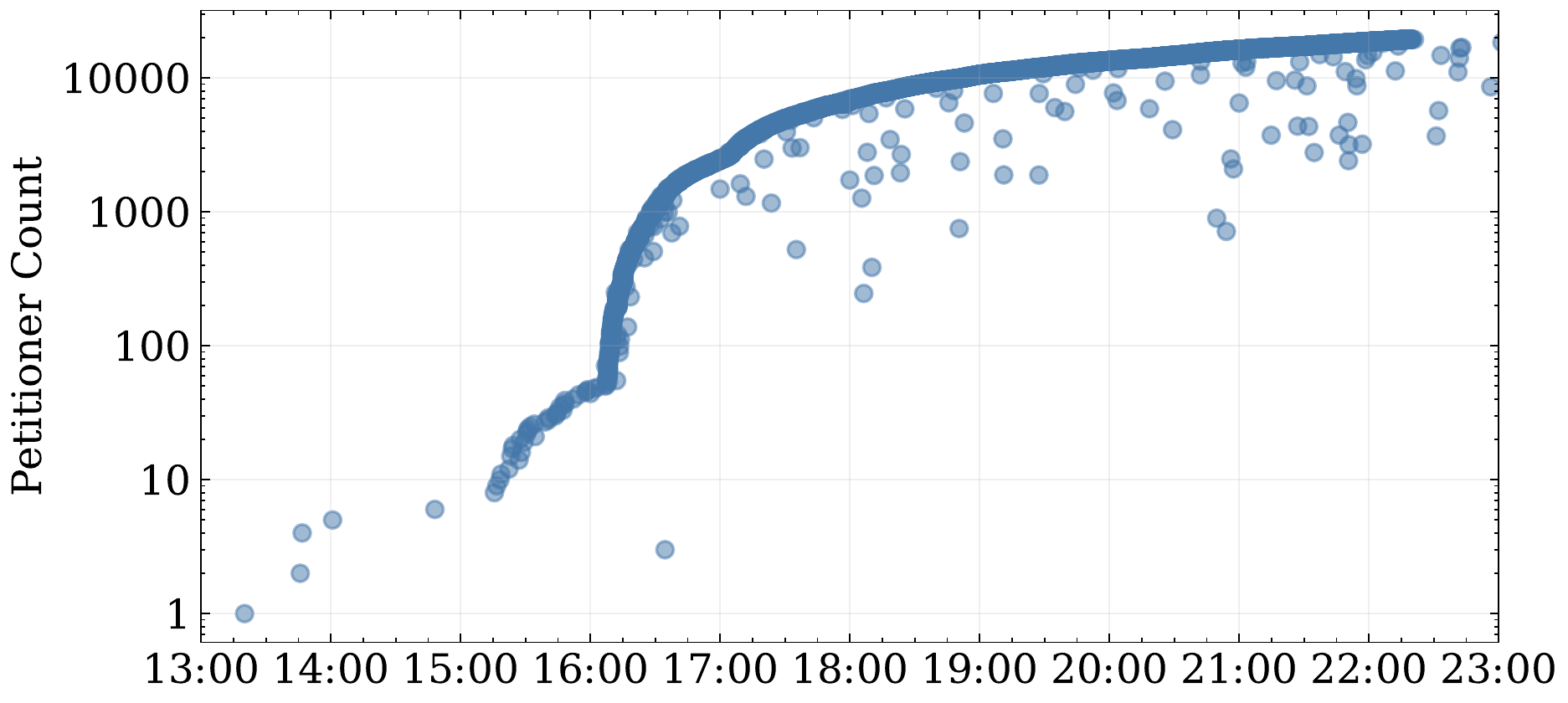}
\centering
\caption{
Number of Weibo petitioners for creation of the help-seeking super-topic on January 29 over time. Outliers may result from network latency.
}
\label{fig:RQ2-petition}
\Description{This figure is fully described in the text.}
\end{figure}

\paragraph{[January 29] Super-Topic Creation: A Self-Organizing Process by Digital Volunteers. }%
Six days after Wuhan Lockdown, the help-seeking super-topic was launched. This was not initiated by Weibo itself, but by a group of users who made petition to the platform. According to Weibo's rules, the creation of a super-topic must be proposed by several users and approved by Weibo.  After joining the petition, users got a fixed-format {post} automatically posted to their timeline, for instance:
``\textit{I am the 234th fan to apply for the creation of `[Super-topic] COVID-19 patient seek help' → http://t.cn/XXX, the more people involved, the greater the chance of it opening! Help speed up the creation by visiting http://t.cn/XXX.}''
Using keyword matching, we located 17,045 petitioners of the help-seeking super-topic within our 100 million records. Figure \ref{fig:RQ2-petition} shows the growth in the number of petitioners (as indicated by their messages, \eg 234) on the entire platform before the super-topic was created. 

The first petitioner appeared after 13:00 on January 29 but did not gain any attention within the next two hours. However, in the three hours following 16:00, the number of petitioners skyrocketed from fewer than 100 to more than 10,000.
By comparing author IDs of petition messages with that of the 8K help requests, we found that 99.5\% of petitioners were not the ones seeking medical assistance. By examining the profiles of the first five petitioners (those who participated before 15:00), we found that two were administrators (see Section \ref{sec:bg-supertopic} for explanation)
of some celebrity-related super-topics, while the remaining three are active users of a fandom. These users may be more familiar with Weibo and its super-topic function. These findings suggest that the initial members of the help-seeking super-topic are more likely to be ``digital volunteers'', 
an emergent group of users who provide crisis-related information and support to affected people \cite{starbird2011voluntweeters, twigg2017emergent}, rather than the groups in need or some official agencies.
\ywj{
While these volunteers did not establish a formal structure like the official volunteer team (as described in Section \ref{sec:bg-supertopic}), they were able to involve earlier in the public response process. It was their creative approach of creating and promoting the help-seeking community that had enabled those in need to form a collective to attract enough attention.
}
Yet it is beyond the scope of this study to analyze these volunteers in detail, but future research can look deeper into their motivations and collaboration process.

Despite the fact that the super-topic community was created with the assistance of many volunteers, help seekers did not widely use it until six days later on February 3, even though the total number of requests by then had reached 1,561.

\paragraph{[February 3 - 4]: Exposure of Super-Topic on Trending List.}
We consider February 3 to 4 as the period when the super-topic really surfaced and began to gain public attention.
At 10:56:02 on February 3, the keyword (``COVID-19 patients seek help'') made it to the trending list, ranking 44/50. Because Weibo's trending index is calculated by capturing keywords or hashtags from popular {posts} \cite{WADMIN2020TT}, this indicates that many people were already discussing the super-topic at that time. In fact, we found that an increasing number of posts on Douban (another social media site with considerable overlapping users with Weibo \cite{zhang2017understanding}) had been referring to the super-topic several hours before it became trending. For examples,
``\textit{Has anyone heard of this \ST ? [link] [screenshot] It makes me feel so helpless and depressed. }''
``\textit{Please follow the \ST! [screenshot] Numerous people are in need. Since this super-topic does not have a administrator, a lot of help-seeking messages are buried by irrelevant posts.}''
However, the trending topic did not last long (less than two minutes according to our data) before it fell off the list. It is unclear whether this is due to the platform's censorship (the administrator of Weibo has stated that ``[the ranking of trending topics] that involve major negative social news or sudden emergencies may be adjusted''\cite{WADMIN2020TT}), or simply because there are not enough audiences following the topic.
However, 
Weibo administrators and authorities 
began taking action on February 4. On that date, Weibo announced a formal partnership with the Wuhan government, and the government reported having assisted 135 cases coming from this site \cite{li2020crisis}. These reactions indicate that the topic is being endorsed and taken up by official organizations. Afterward, the use of the super-topic in help requests started to rise. After February 5, its usage exceeded that of simple {posting} for almost all of the remaining period.

In brief, we found that 
a considerable amount of help requests was made through {posting} before the community achieved public attention. Only after that did help seekers resort to this official channel more frequently.
Nevertheless, many requests continued to be distributed via {posting} function. It is thus unclear to what extent the exposure of the super-topic changed the way people seek help. 
This issue is vital since it would indicate the whether the help-seeking channel was being utilized as adequately as it could be and whether the current settings for communication between help seekers and the emergency responders are effective. 
We further examine this issue in the following subsection.

\subsection{Impact of Super-Topic Exposure on Function Usage }

In order to determine if the super-topic exposure influenced microblogging usage by help seekers, we seek to examine whether people used the super-topic more often after the event, or \textit{if they used the basic posting function significantly less?} In this section, we examine the latter case by modeling the intervention effect of the event on the trend of posted requests.

\paragraph{Controlled Interrupted Time Series Analysis}
Our first consideration is to quantify the intervention effect utilizing an interrupted time series analysis (ITS).

ITS is a powerful quasi-experimental design that evaluates the effectiveness of population-level policies or interventions \cite{bernal2017interrupted}. 
It is also widely used in HCI and social computing research \cite{kizilcec2018social, papakyriakopoulos2021media}.
With segmented regression, ITS can estimate the effect size of an intervention on the outcome regarding the change in level (intercept) and trend (slope). However, we cannot use ITS directly due to other interventions or events that may also affect the outcome in our scenario. The most important reason is that as the medical situation in Hubei improved, all help requests would reduce (see light blue curve after February 9 in Figure \ref{fig:RQ2-time}). 
To address this issue, we adopt controlled interrupted time series analysis (CITS) \cite{lopez2018use}, in which a control series that has not been subjected to the intervention is added to the ITS design.
In our case, the control group is the overall volume of help requests, and the intervention group is the volume of requests made by {posting}. 
While both groups may be affected by external factors, such as improvements in medical conditions, only the intervention group is affected by exposure to the super-topic, as some may begin using this channel in place of the posting function.
We fitted data using the following equation:

\begin{equation} \label{eq1}
\begin{split}
y & = \beta_0 + \beta_1\cdot Time + \beta_2\cdot Phase + \beta_3\cdot Phase\cdot Time \\ & + \beta_4\cdot Group
+ \beta_5\cdot Group\cdot Time + \beta_6\cdot Group \cdot Phase \\ & + \beta_7 \cdot Group \cdot Phase \cdot Time + error
\end{split}
\end{equation}

Where $y$ is the volume of certain group of help requests; 
$Time$ is the time point of data from January 21 to March 15;
$Phase$ is a binary variable representing the intervention (0 before the super-topic exposed on February 3 or 1 after February 4);
$Group$ is a dummy variable that represents the cohort assignment (0 for the control group, or 1 for the intervention group); 
The other variables are all interaction terms. 
The coefficients $\beta$ in equation \ref{eq1} are shown in Table \ref{tab:rq2-CITS}.
As the dependent variable is a count variable, we fit Equation \ref{eq1} by negative binomial regression.
Standard errors are adjusted using the Newey-West method \cite{newey1986simple} to account for autocorrelation and heteroskedasticity \cite{bottomley2019analysing}. Statistical significance was defined as $P < 0.05$.

\begin{table*}[]
\centering
\rowcolors{2}{white}{gray!15}
\resizebox{0.9\textwidth}{!}{
\begin{tabular}{lllll}
\toprule
          & Explanation                                                                                                   & Coefficient (95\% CI) & S.E.  & \textit{P}     \\ \midrule
$\beta_1$ &
\begin{tabular}[c]{@{}l@{}}
The trend of the overall volume of requests prior to intervention\\(\ie the super-topic gained attention between February 3 and February 4)
\end{tabular} &
  0.312 (0.164, 0.460) &
  0.076 &
  0.000  \\
$\beta_2$ & 
\begin{tabular}[c]{@{}l@{}}
The level change in overall volume before and after intervention
\end{tabular}
& 0.409 (-0.934, 1.751)    & 0.685 & 0.550 \\
$\beta_3$ & 
\begin{tabular}[c]{@{}l@{}}
The trend change of overall volume before and after intervention 
\end{tabular}
& -0.449 (-0.599, -0.298)  & 0.077 & 0.000 \\
$\beta_4$ & 
\begin{tabular}[c]{@{}l@{}}
The difference in level between the {posting} volume and overall \\volume before intervention
\end{tabular}
& 0.005 (-1.673, 1.683)    & 0.856 & 0.995 \\
$\beta_5$ &
\begin{tabular}[c]{@{}l@{}}
The difference in trend between the {posting} volume and overall\\volume before intervention 
\end{tabular}
& -0.008 (-0.217, 0.201)   & 0.107 & 0.940 \\
$\beta_6$ &
\begin{tabular}[c]{@{}l@{}}
The difference in level change between the {posting} volume and\\overall volume before and after intervention 
\end{tabular}
& -0.895 (-2.795, 1.004)   & 0.969 & 0.356 \\
$\beta_7$ & 
\begin{tabular}[c]{@{}l@{}}
The difference in trend change between the {posting} volume and\\overall requests before and after intervention
\end{tabular}
& 0.030 (-0.183, 0.244)    & 0.109 & 0.781 \\ \bottomrule
\end{tabular}}
\caption{
Controlled interrupted time-series analysis using negative binomial regression with adjusted Newey-West standard errors. The dependent variable is the volume of help requests.
}
\label{tab:rq2-CITS}
\end{table*}

\begin{figure*}[]
\includegraphics[width=0.8\linewidth]{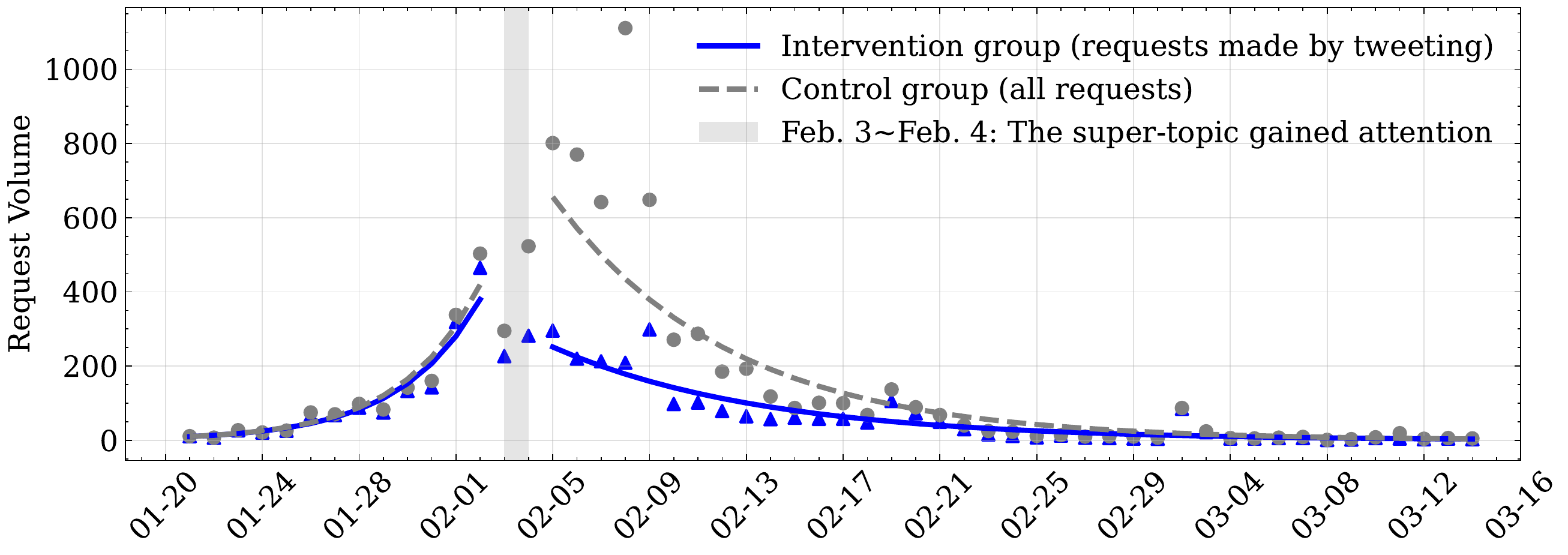}
\centering
\caption{
Controlled Interrupted time series analysis of the volume between help requests made by {posting} (blue triangles) and overall requests (grey dots). The curves represent the fit of the model. The gray area indicates the intervention. 
}
\label{fig:RQ2-transition}
\Description{This figure is fully described in the text.}
\end{figure*}

\paragraph{Result}

Table \ref{tab:rq2-CITS} shows the regression results. There was no significant difference in the mean level ($\beta_4=0.005, CI=[-1.673, 1.683], P=0.995$) and time trend ($\beta_5=-0.008, CI=[-0.217, 0.201], P=0.940$) in volume between requests made by {posting} and overall requests before the super-topic was exposed. Figure \ref{fig:RQ2-transition} also supports this: before February 3, the grey and blue curves are nearly identical. This results ensure that the control group and the intervention group are comparable.
After the exposure of the super-topic, however, the results still show that both the immediate decline ($\beta_6=-0.895, CI=[-2.795, 1.004], P=0.356$) and the trend change ($\beta_7=-0.030, CI=[-0.183, 0.244], P=0.781$) in the volume of requests made by {posting} are insignificant when compared to that of the overall requests.  

The above findings, surprisingly, show that the exposure of the super-topic did not significantly cause people to use the basic {posting} feature less for sending requests. This means even when there is a better channel that may be more likely to receive assistance, affected population kept use the basic {posting} function. 

\section{Discussion}

In this paper, we investigate how people use various microblogging features, especially super-topic, to seek help during the early stage of the COVID-19 crisis and the affordances and limitations of these features. %
Our findings provide several implications on how to facilitate communication between help seekers and emergency responders through better design and inter-organizational coordination.

\subsection{Design Trade-Offs for Crisis Communication Channels 
}

Our work suggests that a centrally managed channel 
can be helpful in responding to urgent requests on social media and reducing some of the inherent limitations in basic social media functionality. 
Weibo's help-seeking super-topic plays a similar role to an ``official hashtag''\cite{acar2011twitter}, directing all stakeholders' attention to one place, making it easier for them to determine the right places for making requests or providing assistance \cite{potts2011tweeting, acar2011twitter}.  
It is also similar to the dedicated Facebook groups that were in service during the 2013 flooding in Europe \cite{kaufhold2014vernetzte}, which utilized specific Facebook pages to mobilize and coordinate resources and offers. But Weibo's help-seeking Super-topic %
additionally implemented a structured posting template \cite{starbird2010tweak} that can relieve the moderators from many of the repetitive and time-consuming tasks associated with information processing. Such dedicated channels can be easily incorporated and operated by relief organizations, enhancing the efficiency of seeking and providing assistance.

However, our analysis of user help-seeking behavior also revealed some challenges in applying emerging channels %
in crisis scenarios. %
First, since the outbreak of the crisis has brought massive information to social media \cite{zarocostas2020fight}, the introduction of new features may be hard to get noticed by all users in a short period, resulting in underused features and a waste of resources.
Second, new tools may come with some learning overhead. In Weibo, for instance, users need to find the Super-topic community entrance and subscribe to it first (not required by other posting features), %
thus complicating the process. Moreover, the index design of super-topic is visually similar to that of hashtags, rendering it difficult for users to distinguish the two channels. As a result, some users posted in a different channel by mistake and failed to enjoy the benefits as intended. %

Therefore, when designing for crisis communication, it is imperative to carefully weigh the trade-offs involved in the development of a new feature or in adapting the existing tools without changing the way they are used.

\subsection{Design Suggestions for Improving Affordances}

To better support the use of microblogging services for emergency assistance, regardless of whether we decide to leverage and improve the existing design or propose new functionalities, the four main limitations outlined in RQ1 Section \ref{limitations} %
should be sufficiently addressed. We offer some design suggestions in this section.

\textit{Request Search Capability.}
 Our machine learning models developed for 
request identification
 demonstrate a potential solution for efficient search of signals from help seekers in microblog streams. Our investigation shows that recent advances in natural language processing, such as transformers \cite{devlin-etal-2019-bert}, can empower a classifier to accurately (in both precision and recall) determine whether a post discloses a request for help simply based on the content of the post. %
 In fact, many researchers have taken an interest in these tools \cite{rogstadius2013crisistracker, ashktorab2014tweedr, imran2014aidr}. AIDR \cite{imran2014aidr} is one example that automatically collects posts and annotates data through crowdsourcing to formulate models that can be used to identify important information during a crisis event (\eg needs and resource donations) quickly. In times of crisis, these tools could be integrated into microblogging platforms to identify users' intentions to ask for help even before they publish the message and allow emergency responders to reach more help seekers faster.

\textit{ Request-Response Tracking Capability.}
 When the original request post is modified or deleted, other related posts which have been propagated by volunteers in the digital space (but do not have a forwarding chain) cannot be updated simultaneously. We observed in our dataset that duplicate postings might occur across different super-topics or hashtags with few modifications by volunteers. Microblogging platforms can employ text-matching techniques such as Simhash \cite{sadowski2007simhash} to identify duplicate posts so as to keep them in syn when updating the feedback on the help seekers' original posts. %

\textit{Ease of Use specially in Urgent Settings.}
 Our findings reveal that the difference between super-topics and hashtags can be confusing for people in terms of the presentation of content and their mechanism. Some digital volunteers 
 have also reported feeling unfamiliar with the use of super-topic \cite{FENG2020SW}. Therefore, rather than relying on users not to make mistakes, the Weibo platform should proactively prevent possible human errors from happening. As an example, when a user is editing a post with a hashtag similar to the one used by official channels for crisis communication, the system could ask if the user wants to submit a help request or proactively suggest the official hashtag/super-topic. The posts could also be automatically uploaded to official channels once the users approve them, instead of asking the users to go through the super-topic guidelines.

\textit{Privacy Protection.}
 In the midst of the crisis, Weibo failed to prioritize privacy protection considerations, and the help-seeking {posts} containing personal information were visible to everyone on Weibo. As soon as the needs are met, however, privacy protection will have to be elevated to a higher priority. In this case, the platform might proactively ask users if they would like to hide these requests and/or hide all the duplicate posts identified via the text matching method discussed above. Another way to protect user privacy is to create emergency channels on the public platform in which the private information is only visible to authorized respondents. For example, on February 5, the People's Daily posted a link to a private questionnaire on Weibo to collect information from help seekers \cite{PDO2020}.%

\subsection{Implication for Communication Between Affected Citizens and Authorities}

Our observation of the help-seeking process on Weibo also suggests that a trusted third party can help coordinate communications between citizens and authorities. Many individuals expect a direct reply from the authorities via the Internet, and they make frequent mentions of official accounts (as well as nonexistent accounts) on social media. These efforts are unlikely to succeed because many barriers prevent emergency managers from responding directly, such as mistrust of social media content and information overload \cite{plotnick2016barriers,haunschild2020sticking}. These factors reduce the likelihood of people obtaining assistance online. Fortunately, the volunteer organization partnered with Weibo could mitigate these obstacles by serving as a connecting liaison between those in need and those who could assist. It is similar to the role of the Virtual Operations Support Team (VOST) \cite{reuter2018fifteen}  in crisis response, which monitors social media channels, gathers information, and reports it to authorities.
The difference is that VOST was set up by emergency managers, while the volunteer %
organization partnered with Weibo was led by citizen volunteers. As moderators will have access to sensitive personal information, such as telephone numbers and addresses, it will be necessary to re-evaluate whether citizen volunteers %
will be suitable to perform this role in the future work.

\subsection{Generalization of Findings}

Despite our focus on Chinese social media, some of our findings may also apply to other microblogging services such as Twitter. 
With the exception of the super-topic, Weibo and Twitter share many similar features, such as hashtags, mentions, and {reposts (retweets)}. In that regard, our evaluation and design recommendations of these similar features might apply to Twitter. In addition, the efficient use of the super-topic for crisis response in Weibo may inspire other microblogging platforms to launch similar channels in face of emergency. 
Nevertheless, some of our findings may not be generalizable to other platforms due to the complex context of the Chinese Internet. For instance, Weibo's policy affecting rankings of trending topics (e.g., controlling negative social events \cite{WADMIN2020TT}) may affect users' awareness of the super-topic, thereby affecting their use of Weibo. 

\subsection{Limitation and Future work}

Our work has several limitations. First, the coverage of the data used in this paper could be impaired by the censorship present on the Chinese Internet. In the super-topic published between February 3 and February 4, some posts were reported to have been deleted by the platform \cite{FENG2020SW}; but some users also indicated that these posts were not help-seeking posts, which explains their removal. In any case, we try to ensure the data coverage by integrating as many publicly available research sources as possible to augment our main dataset that was crawled in real-time. %
Second, the performance of the BERT model was reported on a test set consisting only of posts from the super-topic which may not precisely reflect the overall distribution in the whole dataset. Although some data augmentation techniques are used to improve generalization of the model, we observe some false positives, especially on short texts. This bias is due to the fact that most of our positive samples in the super-topic have relatively long descriptions. Therefore, the classifier makes more errors when identifying the comments part of requests with the quote function related to short text. As a result, we manually screened these posts as described in Section \ref{sec:rq1}, and did not include them in the quantitative analyses. Third, the conclusions drawn from quantitative methods in this paper cannot be explained as causal without natural experiments due to their correlational nature.
Future research could conduct interviews and surveys with these help seekers to verify whether our explanations match the reasons they provide. Fourth, we found that many non-COVID-19 patients sought assistance on Weibo as well. We hope future research will also pay attention to these vulnerable groups.

\subsection{Ethical Consideration}
This study involves utilization of sensitive data on social media, therefore, ethical measures have been taken in an attempt to effectively protect one’s privacy. This research has been approved by the Institutional Review Board (IRB) at the authors’ institution and complaint with the ACM Code of Ethics and Professional Conduct%
. Instances included are being paraphrased, with sensitive information being concealed. Images are being combined from multiple data, with user’s personal information being blurred, ensuring the original content cannot be found on the Internet.

\section{Conclusion}

This paper presents a comprehensive study of the usage of Weibo, a Chinese microblogging platform, by people seeking medical assistance during a public health crisis. From a dataset of more than 100 million {posts} from Weibo, we identified 8K requests for medical assistance from COVID-19 patients or caregivers during the pandemic in China from January to March 2020. 
Through qualitative analysis, we identified several microblogging functions used in the requests and further investigated their limitations in supporting urgent requests. Our findings showed that most help requests were made on an emergent online community (i.e., the help-seeking super-topic) and via basic {posting} (i.e., posting on one's own timeline), the latter of which was heavily used prior to and during the creation of the former but was understudied in previous works. A controlled interrupted time series analysis implied that the exposure of the super-topic on the trending list of Weibo did not significantly influence people's choice of {posting} for making help-seeking requests versus this more effective channel. 
This study provides implications for enhancing the design of microblogging platforms to support help-seeking activities for vulnerable populations during crises and to facilitate crisis relief for emergency responders.

\begin{acks}
We are grateful for the valuable feedback provided by the anonymous reviewers. We thank Qingbo big data for contributing part of the dataset to this study. We also thank Ziyu Wang for his contribution to our brainstorming.
This research was supported in part by ASPIRE League Partnership Seed Fund ASPIRE2021\#3.
\end{acks}

\balance
\bibliographystyle{ACM-Reference-Format}

\end{document}